\documentclass[sigplan, authorversion=true]{acmart}

\usepackage{flushend}
\usepackage{booktabs} 
\usepackage{graphicx}
\usepackage{subcaption}
\usepackage{hyperref}
\usepackage[htt]{hyphenat}

\begin{document}

\title[Optimal DNN Primitive Selection with PBQP]{Optimal DNN Primitive
Selection with\\Partitioned Boolean Quadratic Programming}

\author{Andrew Anderson}
\orcid{1234-5678-9012}
\affiliation{%
  \institution{School of Computer Science and Statistics\\Trinity College Dublin}
  \streetaddress{}
  \city{Dublin}
  \country{Ireland}
}
\email{andersan@cs.tcd.ie}

\author{David Gregg}
\orcid{1234-5678-9012}
\affiliation{%
  \institution{School of Computer Science and Statistics\\Trinity College Dublin}
  \streetaddress{}
  \city{Dublin}
  \country{Ireland}
}
\email{dgregg@cs.tcd.ie}

\renewcommand{\shortauthors}{A. Anderson and D. Gregg}

\begin{abstract}

Deep Neural Networks (DNNs) require very large amounts of computation, and many
different algorithms have been proposed to implement their most expensive
layers, each of which has a large number of variants with different trade-offs
of parallelism, locality, memory footprint, and execution time. In addition,
specific algorithms operate much more efficiently on specialized data layouts.

We state the problem of optimal primitive selection in the presence of data
layout transformations, and show that it is NP-hard by demonstrating an
embedding in the Partitioned Boolean Quadratic Assignment problem (PBQP).
We propose an analytic solution via a PBQP solver, and evaluate our approach
experimentally by optimizing several popular DNNs using a library of more than
70 DNN primitives, on an embedded platform and a general purpose platform. We
show experimentally that significant gains are possible versus the state of the
art vendor libraries by using a principled analytic solution to the problem of
primitive selection in the presence of data layout transformations.

\end{abstract}

%
%




\settopmatter{printacmref=false, printfolios=false}
\setcopyright{none}
\acmISBN{}
\acmDOI{}
\acmYear{2018}
\acmConference[]{preprint}{arxiv.org}{2018}
\let\printcopyright\iffalse
\let\printpermission\iffalse
\maketitle

\section{Motivation}
\label{sec:motivation}

Deep neural networks are among the most successful techniques for processing
image, video, sound and other data arising from real-world sensors. DNNs
require very large amounts of computation which challenge the resource of all
but the most powerful machines.

However, DNNs are most useful when deployed in real-world embedded devices,
which navigate and interact with their surroundings. In these embedded
environments, limits on battery capacity, memory and processing power are
significant constraints.

DNNs consist of a directed graph of ``layers'' that receive raw input data,
and output a transformation or classification of the data. Several different
types of layers are used to implement DNNs, such as activation layers, pooling
layers, convolution layers, and fully-connected layers. In the best-known and
best-performing DNNs, a great majority of the execution time is spent in the
convolution layers.

There are many ways that each layer can be implemented. For example, a common
approach to implementing convolution is to restructure the input data and call
a matrix multiplication routine \cite{Jia:EECS-2014-93}. Other researchers have
used carefully hand-tuned C or assembly routines or specialized domain-specific
algorithms (such as Winograd or FFT convolution) with low asymptotic time complexity.

Given the large space of approaches, it can be difficult to guess which
primitive routine might be best used to instantiate any given layer from a DNN.
With no further information, a simple solution is to simply pick a single
primitive for all layers. However, we show that no single primitive yields best
performance for all layers.

On the contrary, some algorithms perform well across a range of inputs, whereas
others can be quite inefficient on average, but perform extremely well in
particular cases. An additional complication is that input data layouts can
have a large impact on the performance of particular primitives.

However, the outputs of one layer become the inputs of others. Thus data layout
decisions cannot be made in isolation, because the selection of the output
layout of a producer layer determines the input layouts of all connected
consumer layers.

\paragraph{Contributions} In this paper we address the problem of per-layer
primitive selection for deep neural networks to optimize a global goal function
for the whole network. Specifically, we make the following contributions:

\begin{itemize}

\item We demonstrate empirically that different DNN primitives can
  provide very different levels of performance for the same
  layer.

\item We formulate the selection of implementations for convolutional layers in
the presence of data layout transformations as a Partitioned Boolean Quadratic
Programming (PBQP) problem.

\item We demonstrate the effectiveness of our technique using a large
  library of more than 70 DNN primitives operating on a variety of data
  layouts with an off-the-shelf PBQP solver.

\item We demonstrate significant speedups from our approach on desktop and embedded CPUs on several popular
  deep neural networks.

\end{itemize}

\section{Background}
\label{sec:background}

A deep neural network (DNN) consists of a directed graph of \textit{layers}
that receive, process, and output data. Input data enters the graph through an
input layer.  Starting from the input layer(s), each layer of the graph is
executed in topological order. Data flows between layers along directed edges,
which determine the order of execution, similar to data dependences in a basic
block.

The directed graph of layers may be cyclic or acyclic. One well-known class of
acyclic \textit{feedforward} DNNs are Convolutional Neural Networks (CNNs).
CNNs normally accept a large matrix or tensor input, such as an RGB image. The
input layer of the CNN processes the input tensor, and produces one or more
output tensors on its output edge(s).  These outputs trigger the execution of
subsequent layers.

The output of the CNN is commonly a classification, that is, a weighted
distribution of categories --- such as dogs or helicopters --- that the CNN
predicts for the input. Once training is complete, a CNN is stateless; the
output is purely a function of the most recent input and the trained, fixed
internal weights.

The layers within a DNN consist of standard mathematical operators such as
convolution, activation, pooling, and full-connected layers. These standard
operators can be used to build a very large variety of deep learning models. A
DNN can be implemented using a library of \textit{primitive} routines, where
each primitive implements one of each of the types of layer in the DNN. A
single input to a layer is often a large four dimensional tensor. Therefore the
amount of computation performed by each primitive is typically large.

\subsection{Convolution Layers}

The most computationally-intensive layer in many DNNs is the
convolution layer. The convolution layer is based on the idea of
convolving a 2D input of dimensions $H \times W$ with a convolution
kernel of $k \times k$. However, as shown in Figure \ref{fig:mcmk}, in
DNN convolution both the input and kernels have $C$ separate
\textit{channels}. Rather than convolving with just a single kernel,
the $C$-channel input is convolved with $M$ separate $C$-channel
kernels. Thus, DNN convolution operates on a 3D input tensor, a 4D
tensor of kernels, and produces a 3D tensor as output. The total
number of operations in a simple (non-strided) DNN convolution is $O(H
\times W \times C \times k^2 \times M)$.

\begin{figure}[ht]
\centering
\includegraphics[scale=0.15]{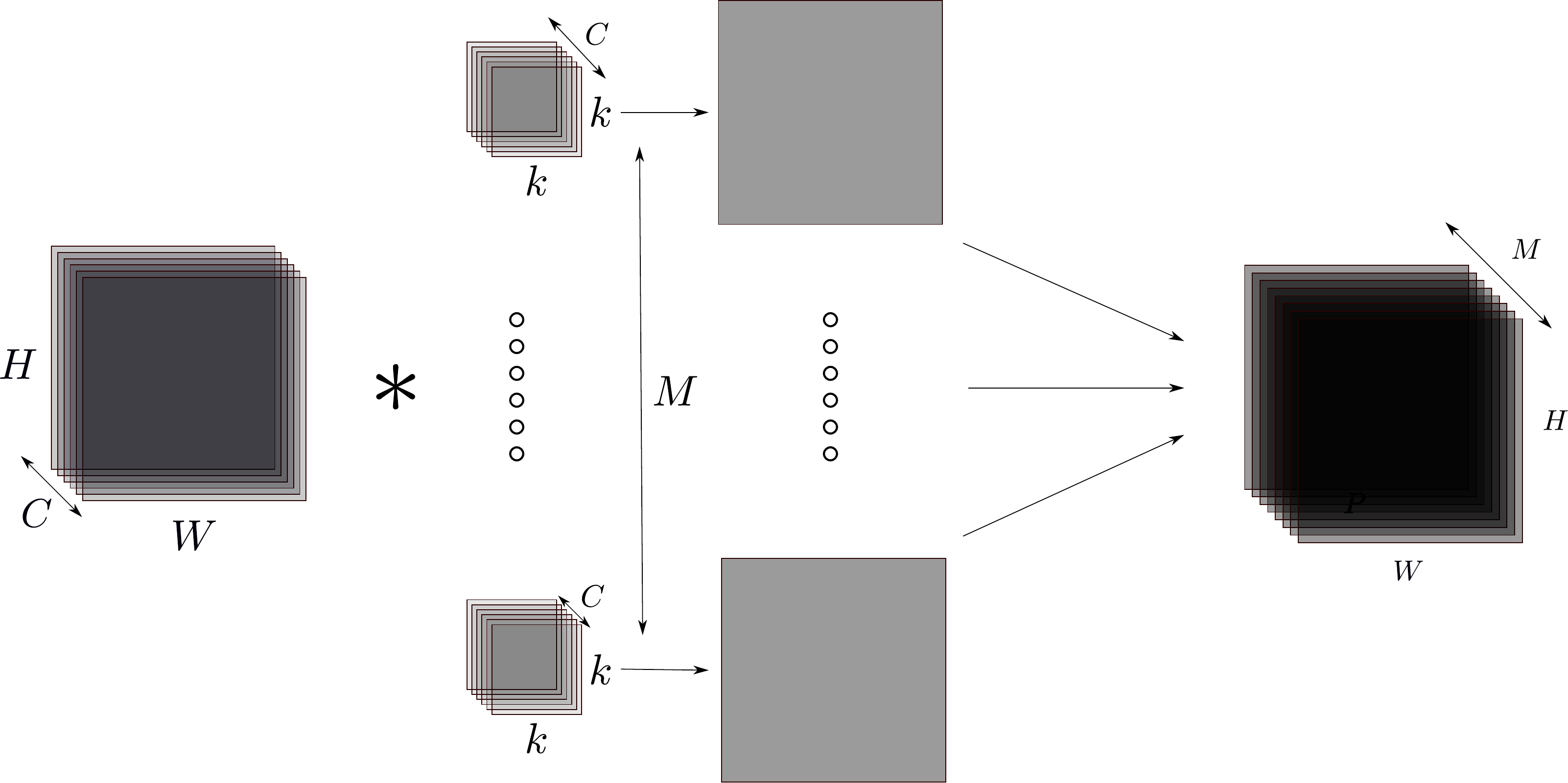}

\caption{DNN convolution of a 3D input tensor composed of $C$ input
feature maps, each of size $H \times W$, with a 4D kernel
composed of $M$ multichannel filters, each with $C$ channels and a $k \times k$
kernel, resulting in $M$ output feature maps, each of size $H \times W$.}

\label{fig:mcmk}
\end{figure}

\subsection{Data Layouts}
\label{sec:data-layouts}
Each primitive takes inputs in a given data layout, performs the operations of
one or more layers, and produces output in a given layout. Some DNN libraries
produce all inputs and outputs in a canonical layout, such as the $NCHW$ data
layout used by Caffe \cite{Jia:EECS-2014-93}. However, forcing each layer to
use the same layout removes the opportunity to customize layouts to particular
operations (such as convolution with large $K$ but small $C$) or for
cross-layer optimizations that exploit specialized layouts.

A simple way to find the cost of implementing any layer with any primitive is
to execute the primitive using sample inputs for the layer. The cost of
execution of most DNN layers depends primarily on the dimensions of the input
rather than on the actual input values. Since the execution time tends to
remain stable regardless of the particular input values, it is possible to
measure with some consistency the execution time of a given primitive
implementing a given layer. However, given a set of execution times for each of
the layers in a network with different primitives, it is not obvious which
primitives to select to implement each layer to maximize overall performance.

In particular, when input data is converted to a special representation, such as a
frequency domain representation, it is often much more efficient to keep the
data in that representation for as long as possible, rather than to convert back and
forth between data representations for each layer in the network.

\section{Primitive Selection}
\label{sec:pbqp}

In this section we consider the problem of selecting a primitive to implement
each layer to yield the lowest cost \emph{instantiation} of a given DNN. By
lowest cost we mean that the sum of the execution times of each of the layers
of the DNN is minimized. This problem may seem no more than selecting the
fastest implementation of each layer, but in fact it is much more difficult.

The parameters of a convolutional layer on which the runtime chiefly depends
are informally well understood --- the work to be done grows with the size and
number of input feature maps, and the size and number of filters, but decreases
as the stride of the convolution increases. We can model a convolutional
\emph{scenario} formally as a 6-tuple $\{C, H, W, \delta, K, M\}$, respectively,
the number of input feature maps, the height of an input feature map, the width
of an input feature map, the stride of the convolution, the radix of the
convolutional filters, and the number of output feature maps.

Note that our formulation does not consider minibatching. Minibatching can
trivially be incorporated by adding a seventh parameter to encode the batch
size. However, our application context is highly latency sensitive, so our
formulation, in practice, considers only a minibatch size of 1.

Input data is provided to a convolutional layer in a 3-dimensional tensor of
size $C \times H \times W$. In the abstract, any \emph{layout} (i.e.
permutation of the order of these dimensions) of the tensor is valid. However,
each primitive operator deals with inputs and outputs in a specific data
layout.

We model primitive operators with a 3-tuple $\{L_{in}, P, L_{out}\}$,
respectively, the input layout, primitive identifier, and output layout, where
the layouts are a permutation of $\{C, H, W\}$. A directed edge from a layer
instantiated with primitive $A$ to another layer instantiated with primitive
$B$ is \emph{legal} iff $L_{out}(A) = L_{in}(B)$. Thus a primitive assignment
also implies a specific layout assignment to the input and output edges of a
DNN layer.

Two incompatible primitives cannot be connected, regardless of the optimality
of such an arrangement. For example, a particular primitive operator that
performs convolution might operate on tensors of 16-bit fixed point data.
Another might operate on 32-bit floating point. If the output data of one
primitive were provided as input to the other, garbage would result.

We combine different incompatible primitives using a \textit{legalization}
phase. The legalization phase inserts additional \textit{data layout
conversion} layers to bisect illegal edges, and legalize an assignment. The
legalizer can then select one or more data layout transformation primitives to
implement the conversion layers.

A problem with selecting data layout transformation primitives is that the number
of supported data layouts may be large. There may not be a separate conversion
primitive connecting every pair of data layouts. This may result in a
\emph{chain} of data layout transformations being required to convert from one
layout to another.

An additional complication that arises from inserting data layout
transformations is that the transformations themselves take time to execute.
The legalization pass may allow two incompatible primitives to work together,
but the cost of the data layout transformations may be so high that the selection
is no longer optimal. Once the cost of data layout transformations is included, the
optimal selection might instead be another selection with fewer or cheaper data
layout conversions.  If the cost of the data layouts is considered only
\textit{after} selection, the solution may be sub-optimal.

Clearly, the problem is more complex than simply picking the fastest legal
primitive for each DNN layer. In fact, as we show in the next section, it is
NP-hard to find the least-cost assignment of primitives to layers in the
presence of data layout transformations.

\subsection{Computing Costs}

We provide a two-stage solution to the primitive selection problem. In the
first stage we compute the cost of converting between each of the supported
data layouts. Note that the set of data transformation routines
between the various pairs of data layouts is not normally complete. Instead we
have a limited set of direct data layout transformation routines between various pairs
of data layouts.

Where there is no direct routine to tranform from data layout $A$ to data
layout $B$, it may be possible to build a chain of transformations between the
layouts. For example, if there exist transformations $A \rightarrow D$, $D
\rightarrow X$ and $X \rightarrow B$, then it is possible to convert from
layout $A$ to $B$ via that chain $A \rightarrow D \rightarrow X \rightarrow B$.

Considering the set of data layouts supported by a DNN library as nodes in a
graph, we can construct a \textit{data-layout transformation (DT) graph}. The
direct data layout transformation routines can be considered directed edges of the
DT graph. If there is a directed path between any pair of nodes in the graph,
then a chain of data layout transformations can be constructed to convert from
one layout to the other. The full set of possible direct and indirect data layout
transformations is then given by the transitive closure of the DT graph.

However, the transitive closure tells us only which pairs of data layouts
\textit{can} be linked by a chain of data layout transformations. We also need
to know the cost of every DT graph path. Thus, we must measure the execution
time of the data layout transformation routines on tensors of the size that
appear at their inputs in the DNN, or else provide a heuristic cost for paths.

In the current paper we measure the execution time of transformations ahead
of time, but simple heuristics might be almost as effective.

To find the least-cost chain of data layout transformations between a pair of data
layouts we need to find the shortest path between the corresponding pair of DT
graph nodes. Rather than computing the shortest path between each pair of nodes
each time we need it, we instead compute the all-pairs shortest path for the DT
graph ahead of time. Where no path exists between a pair of nodes, the cost of
the data layout transformation is infinite. This gives us the (transitive) data
layout transformation costs for the transfer of data between \textit{layers} of
the original DNN graph.

In addition to data layout transformation costs, we also compute the cost of
implementing each of the primitive functions that implement \textit{layers} of
the DNN graph. Each layer might be implemented by one of many different
primitives. For example, in our DNN library, we provide implementations of six
major \emph{families} of convolution algorithm. Each algorithm has many
variants with distinct performance characteristics, resulting in over 70
different primitive routines that implement DNN convolution.

To estimate the cost of a specific assignment of a primitive to a DNN layer, we
profile the execution time of the primitive operating on tensors of the size
used in the layer. Note that the dimensions of all inputs to DNN layers are
known statically, and the control flow in most layers is sensitive only to the
size of inputs, not the specific values of the input data. Therefore,
statically-measured execution times on random input of the appropriate size
give a very good estimate of the actual execution time.

\subsection{The Optimization Problem}

Given a DNN graph $G$, we compute the instance cost (execution time) for all
convolutional scenarios $S \in G$ under every available primitive $P_0...P_n$
to form the product space $S \times P$. Since each edge cost is determined by
the \textit{pair of assignments of primitives to nodes in each of the nodes
linked by the edge}, we must then compute all paths $D$ in the DT graph implied
by each \emph{assignment}, $I$, of the nodes of $G$, drawn from points in this
space, and compute the instance cost (execution time) for these also. The task
before us is then to find the point in the product space $S \times P \times D$
of instantiations of $G$ which minimizes the total cost.

Constructed in this way, we can map the problem of layer selection in the
presence of data layout transformations to a well-known existing optimization
problem --- PBQP.

\subsection{Partitioned Boolean Quadratic Programming}

The partitioned Boolean quadratic programming (PBQP) problem is an assignment
problem that can be visualized as a graph. For each node there are several
possible assignments, each with a known cost. In addition, there is a second
set of costs associated with edges in the graph.  The cost of edges is a table
indexed by the pair of assignments of the two nodes linked by the edge. PBQP
has been used to model a number of problems in compiler optimization such as
register allocation for architectures with irregular instruction sets
\cite{Scholz:2002}, and instruction selection on DAGs \cite{Eckstein2003}.

Given a set of layers and a set of primitives, we attempt to assign the layers
to primitives to minimize the total cost. The cost of assigning a layer to a
primitive is the execution time of the layer implemented with that primitive.
Figure \ref{fig:pbqp-nodes} shows a simple graph representation of a PBQP
instance.

\begin{figure}[h]
    \centering
    \begin{subfigure}[b]{0.45\linewidth}
    \centering
    \includegraphics[scale=0.45]{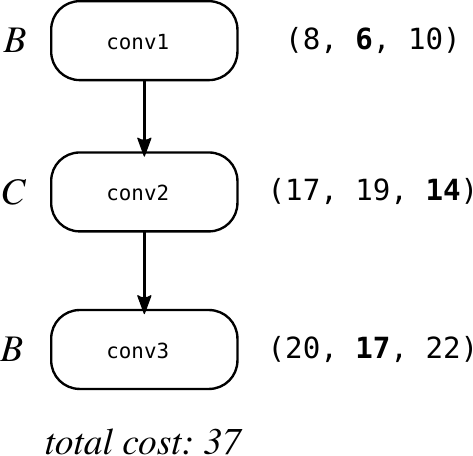}
    \caption{Node costs only}
    \label{fig:pbqp-nodes}
    \end{subfigure}
    ~
    \begin{subfigure}[b]{0.45\linewidth}
    \centering
    \includegraphics[scale=0.45]{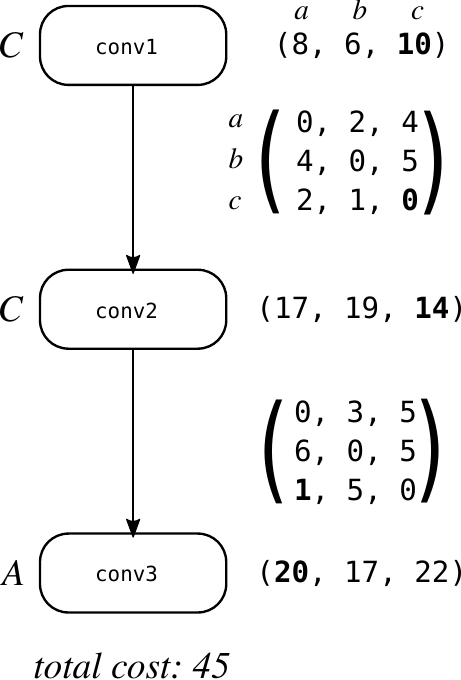}
    \caption{Node and edge costs}
    \label{fig:pbqp-with-edges}
    \end{subfigure}
    \caption{Example of simple linear PBQP problem. Optimal assignments are indicated by the letter left of the node.}\label{fig:pbqp-fig}
\end{figure}

Each layer can be implemented by any one of three different
primitives: $A$, $B$ or $C$. Each of the three primitives uses a different
algorithm, and therefore has a different cost to implement the layer.

Given the cost of implementing the nodes with each of the three primitives in
Figure \ref{fig:pbqp-nodes}, the optimal primitive selection for each node is
clear. The lowest cost selection for each of the three layers is $B$, $C$ and
$B$ respectively.

In addition to the cost table for each individual layer, the PBQP formulation
allows a second cost to be specified for edges. In our simple example, an
additional cost matrix can be associated with each edge in the graph. The cost
matrix for an edge represents the costs associated with all pairs of selections
for the two nodes connected by the edge.

Figure \ref{fig:pbqp-with-edges} shows a cost table for the edge that
transitions from \texttt{conv1} and \texttt{conv2}. For example, if primitive
$A$ is selected for \texttt{conv1} and $A$ is also selected for \texttt{conv2},
then the edge cost of transitioning between \texttt{conv1} and \texttt{conv2}
is zero. This zero cost is the result of using the same data layout for both
\texttt{conv1} and \texttt{conv2}. In contrast, in this example we assume that
primitives $A$, $B$ and $C$ operate on different data layouts, and there is
therefore a data layout transformation cost which arises when we transition
between different primitives in connected layers.

In Figure \ref{fig:pbqp-with-edges} primitive $B$ has the lowest cost for layer
\texttt{conv1}. However, when we consider data layout transformation costs, primitive
$B$ is no longer the optimal selection for layer \texttt{conv1}. For layer
\texttt{conv2}, primitive $C$ is so much faster than the other two choices.
However, the cost of transitioning from $B$ at layer \texttt{conv1} and $C$ at
layer \texttt{conv2} is so high that $B$ is not the optimal selection for layer
\texttt{conv1}. The edge cost is large because the two primitives use different
data layouts, and therefore a data layout transformation must be inserted
between the two primitives. The additional cost that arises from pairs of
assignment choices on the graph edges makes PBQP different from other similar
problems, such as the \textit{quadratic assignment problem}
\cite{Koopmans:1957}.

\begin{figure}[ht]
        \centering
        \includegraphics[scale=0.4]{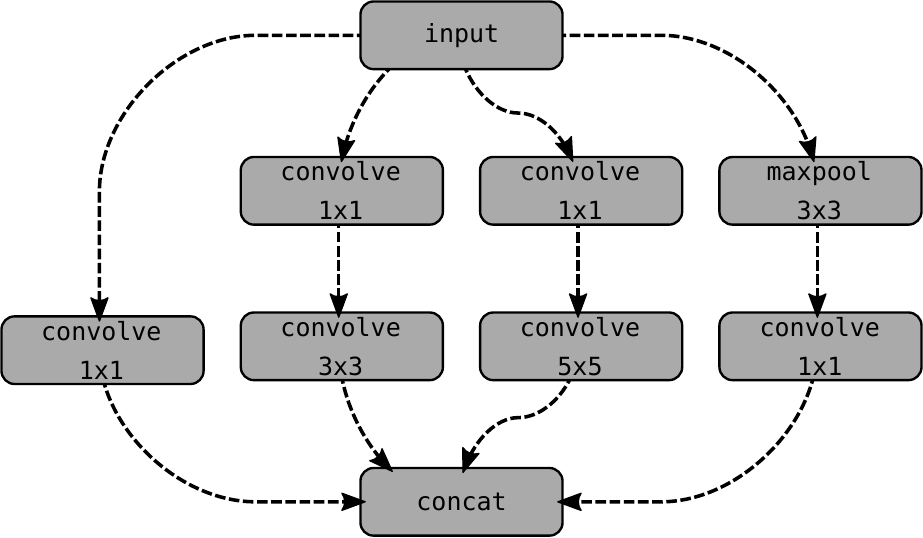}
        \caption{Inception Module}
        \label{fig:inception}
\end{figure}

In the simple example in Figure \ref{fig:pbqp-fig} it is feasible to solve the
selection problem relatively simply even in the presence of edge costs. This is
partly because the example is small, and partly because the graph in the
example is purely linear.  However, data layout transformation costs on graph
edges are much more problematic in DAG-shaped DNNs. Figure \ref{fig:inception}
shows an example of a DAG sub-graph from the GoogleNet
DNN~\cite{DBLP:conf/cvpr/SzegedyLJSRAEVR15}.

Where a layer has multiple direct successors and/or predecessors, the same data
layout may not be optimal for all. Choosing one layout over another may limit
the choice of primitives that can be used in successor/predecessor layers, or
may require the insertion of expensive data layout transformations.  PBQP
allows us to model these complicated DAG costs within an abstract optimization
problem, and to find solutions with an off-the-shelf PBQP solver.

\section{DNN Convolution Algorithms}
\label{sec:algorithms}

A wide variety of convolution algorithms have been proposed over many years,
each with their own strengths and weaknesses \cite{Blahut:2010}. Researchers in
signal processing often distinguish between \textit{direct} and
\textit{indirect} convolution methods.

Direct methods compute the convolution as a simple sum of products, as is found
in a standard textbook definition of convolution. On the other hand so-called
\textit{fast convolution algorithms} reduce the computational complexity by
transforming the input to another form before performing the convolution. In
this section we describe several different classes of convolution algorithms
that are commonly used to implement DNNs.

\noindent $\bullet$ The \emph{direct-loop} family of convolution algorithms
perform multi-channel multi-kernel convolution using a simple six-deep loop
nest. There are many variants of this loop nest with different reorderings,
tilings, and schedules to improve execution time, vectorization, and spatial
and temporal locality of data access.

The \emph{sum-of-single-channels} algorithm is a member of the
\emph{direct-loop} family of methods. It uses the loop ordering $M
\times C \times H \times W \times K \times K$.

\noindent $\bullet$ The \emph{im2} family of convolution algorithms are
variants of the well-known \emph{im2col} approach~\cite{Jia:EECS-2014-93}.
These convolutions first construct a Toeplitz matrix from the input image, and
convolve this with the kernel using a single call to the BLAS GEMM routine.

\noindent $\bullet$The \emph{kn2} family of low-memory GEMM-based convolution
algorithms are presented by Vasudevan et
al.~\cite{DBLP:conf/asap/VasudevanAG17,VasudevanAKG:2017}. This family of
approaches does not construct a Toeplitz matrix, and instead computes
convolution as the sum of several matrix multiplications.  We use
variants of the \emph{kn2} family that compute the sum of GEMMs as an
accumulation and achieve good execution times with low additional memory.

\noindent $\bullet$ The \emph{Winograd} family of methods use the Winograd
algorithm for convolution with a theoretically optimal number of
multiplications~\cite{Blahut:2010}. Unlike the \textit{direct-loop, im2} and
\textit{kn2} approaches, all of which perform the same number of floating point
operations, \textit{winograd} is a ``fast'' algorithm by the signal processing
definition, which greatly reduces the number of operations. We implemented the
Winograd algorithm for scenarios with $K = 3$ and $K = 5$.

\noindent $\bullet$ The \emph{fft} family of methods perform FFT convolution
via the convolution theorem, by first computing the Fourier transform of the
input image and the kernel, applying a pointwise multiplication, and then
computing the inverse Fourier transform of the resulting matrix to produce the
output. Our \textit{fft} implementations compute 2D convolution as a sum of 1D
FFT convolutions, which requires less space than 2D FFT convolution at the cost
of more operations.

\begin{table}[h]
  \caption{Strengths and weaknesses of different convolution algorithms commonly used in DNN implementations}
  \label{tab:algorithms}
\begin{tabular}{|l|c|c|c|c|l|}
 \hline
 Algorithm & Time & Memory & Strided & Bad cases \\
 \hline
 direct loop & - - & ++   & ++  & Non-strided \\
 im2         & +   & - -  & ++  & Large image \\
 kn2         & +   & +    & - - & Few channels \\
 Winograd    & ++  & -    & -   & Unpredictable \\
 FFT         &     & -    & +   & Small kernel \\
 \hline
\end{tabular}
\end{table}

Table \ref{tab:algorithms} gives a brief overview of some of the strengths and
weakneses of these major approaches. In principle it should be possible to
optimize the simple loop nest to achieve high performance, but in practice such
loop nests are more often very slow. The Winograd algorithm \textit{can} be
very fast for $K = 3$ and $K = 5$ convolutions, but equally there are cases
where the \textit{im2} and \textit{kn2} approaches are faster. We have found it
very difficult to predict the cases where Winograd will be faster without
measuring the performance of the layers. The \textit{kn2} approach is fast and
requires little memory, but it cannot be used to efficiently implement strided
convolutions. On the other hand, our FFT algorithm is only sometimes faster
than other approaches, but in those rare cases it can give significant
speedups.

\paragraph{Real-World Solutions}

To demonstrate our formulation in practice, Figure \ref{fig:selection} shows
two primitive selections made by the formulation. Figure
\ref{fig:selection} shows the convolution layers of
AlexNet~\cite{DBLP:conf/nips/KrizhevskySH12}. The figure shows the
selection of primitives that our approach makes on ARM Cortex A-57 and Intel
Core i5-4570 processors.

\begin{figure}[ht]
  \centering
  \includegraphics[width=\linewidth]{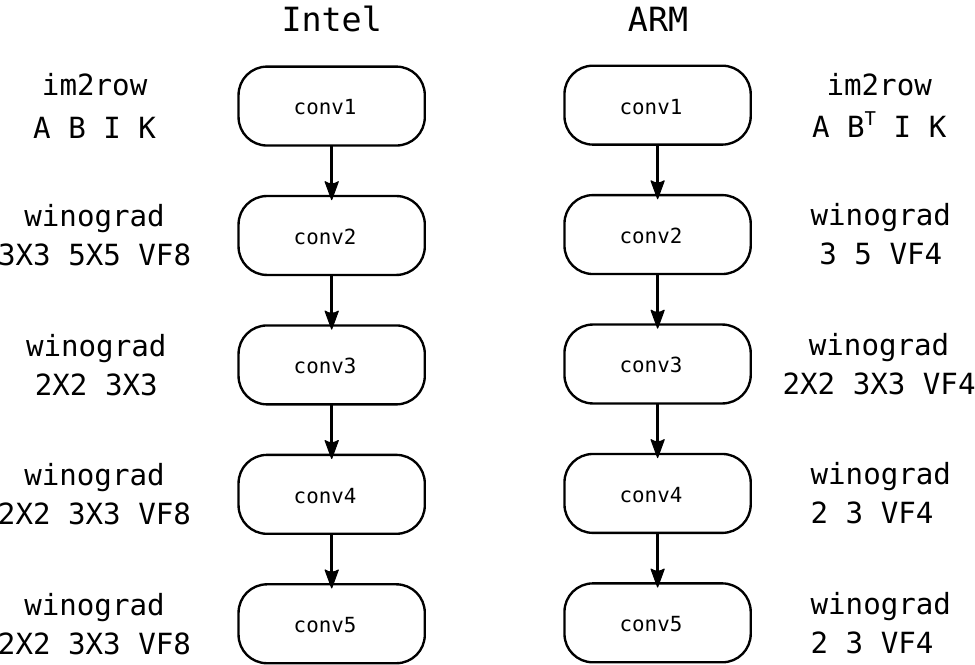}

  \caption{PBQP selections for \textbf{multithreaded execution} of
  AlexNet on ARM Cortex A-57 and Intel Core i5-4570.}

  \label{fig:selection}
\end{figure}

The primitive selections on both processors are interesting in both their
similarities and differences. Recall that we select from a library of around 70
primitive functions which implement DNN convolution, and therefore the range of
possible selections is large.

The \texttt{conv1} layer of AlexNet consists of a $K = 11, \delta = 4$
convolution. The layers selected for both ARM and Intel processors are
\textit{im2} layers, with a row-oriented layout. The only difference between
the two is that the layer selected for ARM passes the kernel matrix to the GEMM
matrix multiplication call as a transposed matrix. There is no good way to
select between these very similar primitives except by profiling the code to
see which is faster on a particular target architecture.

The remaining layer selections are all instances of the Winograd convolution
algorithm for both target architectures. However, there is a striking
difference in the selections for the ARM and Intel processors. The selection
for the Intel processor consists entirely of primitives that implement
two-dimensional Winograd convolution. This approach requires significant
memory, but minimizes the number of executed operations.

In contrast, a majority of the Winograd selections for the ARM processor are
one-dimensional Winograd convolutions. One dimensional Winograd convolutions
require more floating point operations than their 2D counterparts, because the
2D operation must be constructed from the sum of a number of 1D Winograd
convolutions. However, the 1D algorithm requires less memory. Given that the
ARM processor has much smaller caches than the Intel processor, it seems that
the lower memory requirements of 1D Winograd makes it faster on the ARM
processor. As a result, of the four Winograd primtives that are selected to
implement convolution layers on the ARM processor, three are 1D Winograd
convolution algorithms.

Note also that the optimizer selects the 8-way vectorized implementations of
Winograd on the Intel processor, which has Intel's 8-way FP32 AVX2 vector
extensions, while on the ARM processor, which has ARM's 4-way FP32 NEON vector
extensions, it also selects the appropriate architectural variant.

A strength of our approach is that we can easily capture these fine
architectural differences with layerwise profiling, while keeping the optimizer
free from platform-specific special cases. Layerwise profiling need only be run
once per hardware platform per DNN model. The resulting cost tables are tiny
compared to the weight data required for most DNN models, making it feasible to
produce these cost tables before deployment, and ship them with the trained
model to maximise inference performance in situ.

\section{Experimental Evaluation}
\label{sec:experiments}

To validate our primitive selection approach, we performed whole-network
benchmarking with several popular CNN architectures on multiple hardware
platforms. We also ran these CNN architectures using some state-of-the-art
software frameworks, to demonstrate the benefit of our global approach to
optimization of DNN inference. This section presents the results of our
evaluation, with discussion.

\subsection{Experimental Setup}

We evaluated our approach on an Intel Haswell CPU, specifically the Intel Core
i5-4570, and on an ARM Cortex A-57 CPU. The Cortex A-57 is used in NVIDIA's
Tegra X1 platform for embedded and automotive development.

On the ARM platform, we used the popular BVLC Caffe
framework~\cite{jia2014caffe}, which is accelerated by the high-performance
OpenBLAS library under the hood. On the Intel platform, there is a vendor
library available with an optimizing code generator which targets DNN
inference: MKL-DNN~\cite{mkldnn}. The BVLC Caffe framework also runs on this platform.

We used Caffe version 1.0 on all platforms, and on the Intel platform, we used
MKL-DNN version 0.10. We used OpenBLAS release 0.2.20 on all platforms. Both
Caffe and our own primitive library use OpenBLAS internally to perform the
matrix multiplications at the heart of DNN convolution, so using the same
underlying OpenBLAS library ensures we are competing on an equal footing.

We kept all external software dependencies in sync between each platform. The
C++ compiler we used was GCC stable release 7.2. All benchmark executables were
compiled using \texttt{-march=native} \texttt{-std=c++14} \texttt{-O3}. The
NVIDIA Tegra X1 board we used for the Cortex-A57 benchmarking was running
NVIDIA JetPack 3.1.

\subsection{Methodology}

For performance evaluation of our approach, we used three popular network
architectures: AlexNet~\cite{DBLP:conf/nips/KrizhevskySH12},
VGG~\cite{DBLP:journals/corr/SimonyanZ14a}, and
GoogleNet~\cite{DBLP:conf/cvpr/SzegedyLJSRAEVR15}.

Each of these network architectures has a public model corresponding to the
publication made available by the authors, or via the BVLC Caffe Model Zoo, a
public repository of trained DNN architectures. We used these public versions
of the network architectures for our experimentation.

Note that Caffe models other than VGG-D and VGG-E have not been released by the
authors~\cite{DBLP:journals/corr/SimonyanZ14a}, so we cannot present Caffe
results for these models. However, we present results obtained from a faithful
reconstruction of the models by hand, exactly
following~\cite{DBLP:journals/corr/SimonyanZ14a}.

Starting from the published network models, we extracted all convolutional
scenarios in the graph, performed the profiling to gather cost data, and
constructed the PBQP query for the minimum cost instantiation. We mapped the
solution to code with a simple code generator which emitted calls to primitive
operations in our library.

Note that we modelled \textbf{convolutional layers only} in the DNN graph for
optimization. All other layers were represented in our formulation as dummy
nodes, accepting any input and output layouts, and having zero cost. Since
convolution accounts for a very large proportion of the execution time of a
DNN, we hoped that this simplifying assumption would reduce the size of our
optimization queries, while not damaging the quality of the solution.

We used the same code generator with the example code which ships with Intel's
MKL-DNN library to obtain executable instances of the models using
MKL-DNN\footnote{Intel's state-of-the-art MKL-DNN framework~\cite{mkldnn}
contains an optimizing JIT compiler for DNN convolution. The framework natively
targets Intel vector extensions, and underpins projects such as Intel
Caffe~\cite{intelcaffe}.}. For Caffe, we executed published network models
directly using \texttt{caffe time}. Every whole-network benchmark was run for
five iterations, and the mean execution time for one forward pass was used to
calculate speedups.

We performed separate single-threaded and multi-threaded cost modelling, and
solved and built independently two versions of every benchmark program. The
multi-threaded benchmarks were run using all cores available on the machine.
Both of the test systems had four CPU cores. On our graphs, all bars represent
a speedup over a common baseline. For this baseline, all convolutions in the
network are performed using the textbook sum-of-single-channels algorithm, with
single-threaded execution.

\subsection{Data Layouts Used}

Published algorithms for Winograd convolution, \textit{im2} and \textit{kn2}
family convolutions, and direct convolutions tend to use one of three main data
layouts for inputs and outputs: $C \times H \times W$, $H \times C \times W$,
and $H \times W \times C$. The latter two layouts correspond to transpositions
or blockings of the simple data layout used by direct convolutions ($C \times H
\times W$). Note that the \emph{same} data layout is not always used for both
input and output by a primitive operation.

All layers and all primitives used to instantiate them in every network in our
benchmarking operate on 32-bit single-precision floating point data.

\subsection{Optimization Overheads}

We used the PBQP solver of Scholz et al.~\cite{Hames2006} to solve our
optimization queries. Solving the PBQP optimization query took less than one
second for each of the networks we experimented with, using our Intel
experimental machine. In each case, the solver reported that the optimal
solution was found.

\subsection{Interpretation of Results}

Our presentation of results from whole-network benchmarking labels results with
the name of the family of convolution algorithms which was used to produce the
result. For each family of algorithms, we construct the test network by picking
the fastest variant of that family to replace the sum-of-single-channels
algorithm for each individual convolution in the network, if the replacement
is, in fact, faster than sum-of-single-channels for that convolutional scenario
in benchmarking. From this we obtain bars for each family of convolution
algorithms from Section~\ref{sec:algorithms}: \emph{direct, im2, kn2, winograd,
fft}.

We also tested the simple strategy outlined in Section~\ref{sec:data-layouts}
which eliminates all data layout tranformations by choosing a canonical layout
for tensors, and keeping all data in this format. This is the \emph{local
optimal} bar on the graphs. For our experiments, we used the default Caffe
layout, $C \times H \times W$, as the canonical layout.

Absolute timings for the subset of networks which run on both experimental
architectures are presented in Tables~\ref{tab:timings-haswell}
and~\ref{tab:timings-arm}.

\subsection{Experimental Results on Core i5-4570}

Figures~\ref{fig:results-x86-single-threaded} and
~\ref{fig:results-x86-multi-threaded} present the results of our whole-network
evaluation on the Intel Haswell platform. The solution found by the PBQP
formulation for single threaded execution of a DNN model is competitive with
the optimized vendor library, even outperforming it in some cases, such as on
the VGG-B, VGG-C, and VGG-E DNN models. However, it is in multithreaded
execution where the PBQP approach really shines. The PBQP solver finds
excellent solutions for all DNN models, outperforming the vendor library by as
much as a factor of 2x in the case of the VGG-E model
(Figure~\ref{fig:results-x86-multi-threaded}).

\begin{figure*}[ht]
        \centering
        \includegraphics[width=\linewidth]{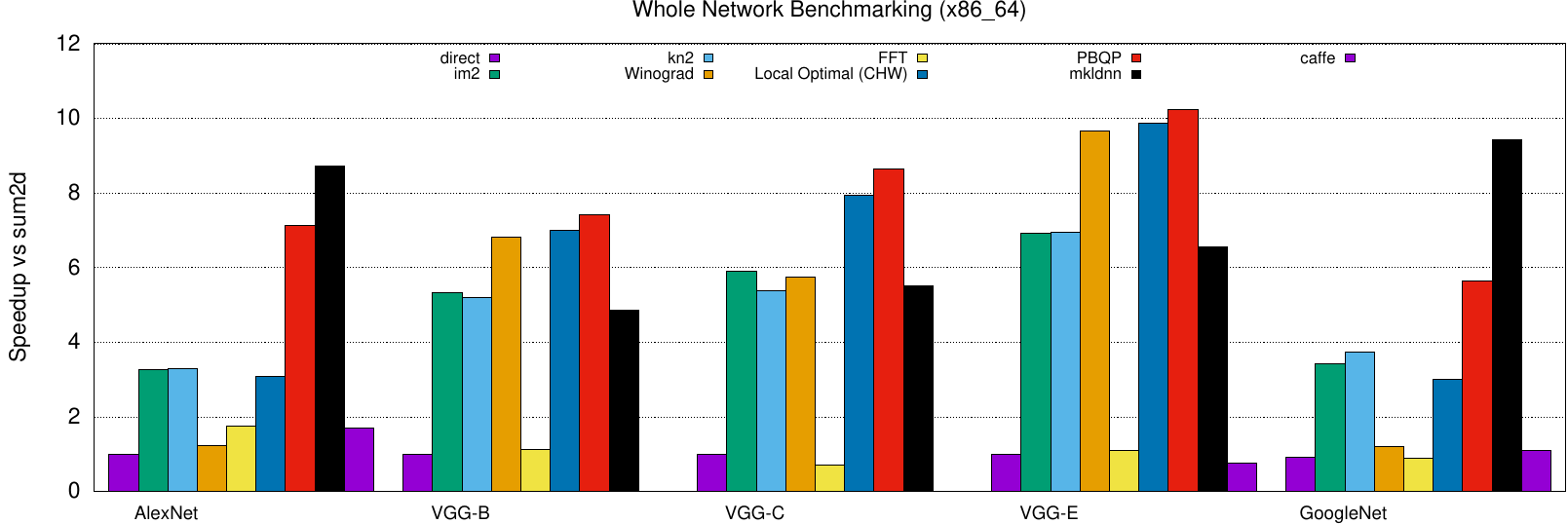}
        \caption{\textbf{Single-Threaded} Comparison of approaches with PBQP selection on \textbf{Intel Haswell}}
        \label{fig:results-x86-single-threaded}
\end{figure*}

\begin{figure*}[ht]
        \centering
        \includegraphics[width=\linewidth]{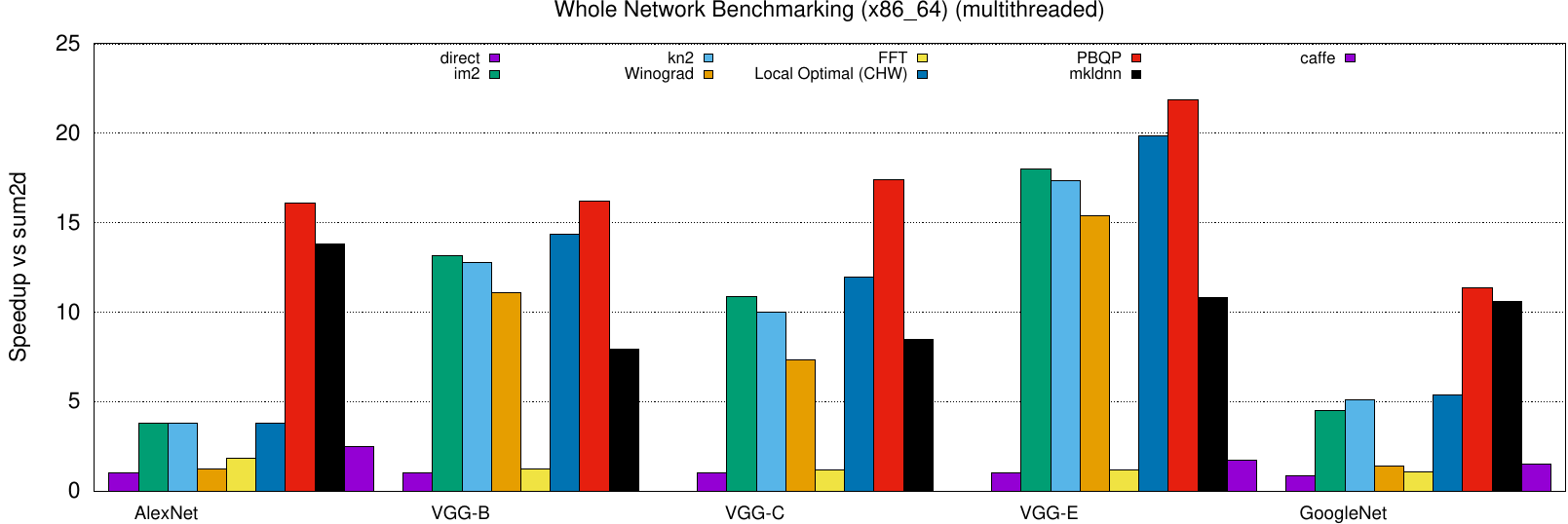}
        \caption{\textbf{Multi-Threaded} Comparison of approaches with PBQP selection on \textbf{Intel Haswell}}
        \label{fig:results-x86-multi-threaded}
\end{figure*}

\subsection{Experimental Results on Cortex-A57}

Figure~\ref{fig:results-a57} presents the results of our whole-network
evaluation on the ARM Cortex-A57 platform. Note that the VGG-B, VGG-C, and
VGG-E DNN models are too large to fit on this platform. However we were able to
reliably execute the AlexNet and GoogleNet DNN models.

Although less pronounced than on Intel Haswell, we still see a very significant
speedup (up to 7x) from our approach versus Caffe on the Cortex-A57
(Figure~\ref{fig:results-a57-multi-threaded}).

\begin{figure*}
\begin{subfigure}{0.5\linewidth}
\centering
\includegraphics[width=\linewidth]{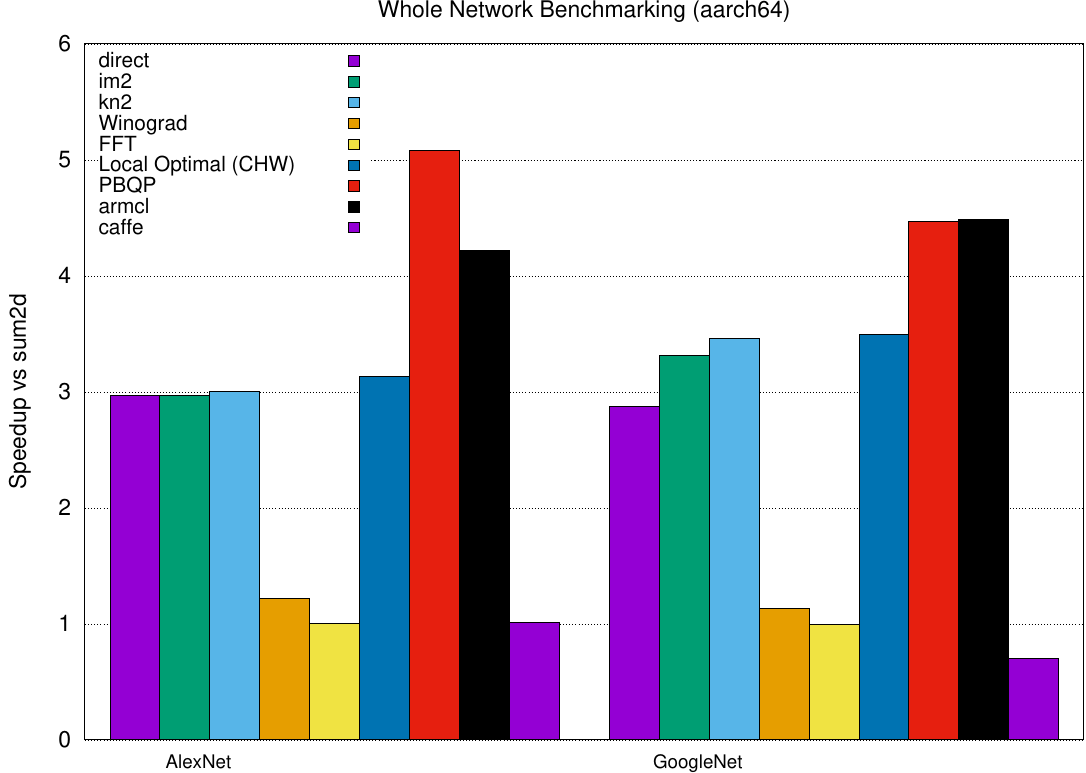}
\caption{\textbf{Single-Threaded} Comparison of approaches with \\PBQP selection on \textbf{ARM Cortex-A57}}
\label{fig:results-a57-single-threaded}
\end{subfigure}%
\begin{subfigure}{0.5\linewidth}
\centering
\includegraphics[width=\linewidth]{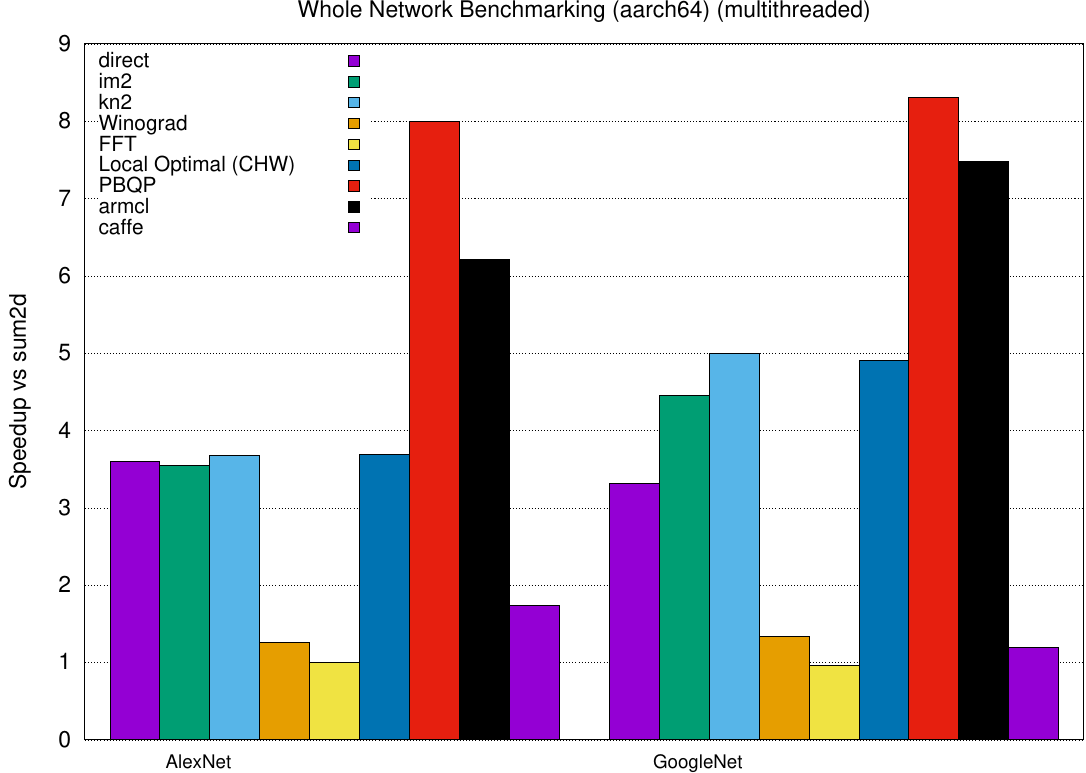}
\caption{\textbf{Multi-Threaded} Comparison of approaches with \\PBQP selection on \textbf{ARM Cortex-A57}}
\label{fig:results-a57-multi-threaded}
\end{subfigure}
\caption{Whole-network benchmarking results on \textbf{ARM Cortex-A57}}
\label{fig:results-a57}
\end{figure*}

\subsection{Experimental Trends}

Our results strongly support the position that there is no one convolution
algorithm which excels in every scenario. For example, Winograd convolution is
supremely effective in VGG family DNN models, because they are entirely
composed of $K = 3$ convolution layers. However, Winograd by itself
is among the slowest approaches for the AlexNet and GoogleNet models
(Figure~\ref{fig:results-x86-multi-threaded}).

We also see an experimental validation of the importance of modelling the cost
of data layout transformations. Recall that our strategy for non-PBQP bars is
to replace the \texttt{SUM2D} algorithm only if the replacement has a
\emph{smaller} execution time. Figure~\ref{fig:results-a57-single-threaded}
shows that for GoogleNet, this strategy leads to a net \emph{slowdown} for the
family of direct loop-based convolutions --- even though at all points the
\emph{faster} loop was chosen, the legalizing data layout transformations are
so expensive that the baseline \texttt{SUM2D} instantiation of the network is
faster.

The effect is even larger considering the case of Winograd convolution for
AlexNet on Intel Haswell (Figure~\ref{fig:results-x86-multi-threaded}). Recall
from Figure~\ref{fig:selection} that the PBQP optimal selection on this
platform uses a Winograd variant for 4 of the 5 convolutions in AlexNet. Yet
simply selecting the fastest Winograd variant ignoring data layout transformation
costs yields an instantiation that performs only marginally better than the
baseline \texttt{SUM2D} network instantiation.

\begin{table}[t]
  \caption{Single inference time on Intel Core i5-4570 (ms) with (S)ingle-threaded and (M)ulti-threaded execution.}
  \label{tab:timings-haswell}
\begin{tabular}{|l|c|c|c|c|c|}
 \hline
 Network & \texttt{SUM2D} & \texttt{L.OPT} & \texttt{PBQP} & \texttt{CAFFE} \\
 \hline
 (S) AlexNet & 711.75 & 231.75 & 100 & 419.565 \\
 (S) GoogleNet & 1401 & 465.25 & 249 & 1267.07 \\
 \hline
 (M) AlexNet & 712.25 & 186 & 44.25 & 286.518 \\
 (M) GoogleNet & 1400.25 & 261.5 & 123.5 & 919.196 \\
 \hline
\end{tabular}
\end{table}

\begin{table}[t]
  \caption{Single inference time on ARM Cortex A-57 (ms) with (S)ingle-threaded and (M)ulti-threaded execution.}
  \label{tab:timings-arm}
\begin{tabular}{|l|c|c|c|c|c|}
 \hline
 Network & \texttt{SUM2D} & \texttt{L.OPT} & \texttt{PBQP} & \texttt{CAFFE} \\
 \hline
 (S) AlexNet   & 2369.5   & 744.25  & 461   & 2341.09 \\
 (S) GoogleNet & 4544.75  & 1695.25 & 1025  & 5782.4  \\
 \hline
 (M) AlexNet & 2432.5 & 639.25 & 294 & 1342.62 \\
 (M) GoogleNet & 4509.75 & 919.25 & 547.5 & 3707.91 \\
 \hline
\end{tabular}
\end{table}

\section{Discussion}
\label{sec:discussion}

Our experimental results demonstrate that our PBQP formulation is
extremely effective at selecting the optimal primitives to implement a
DNN. In this section we briefly consider some alternative strategies
to the same problem, and discuss their strengths and weaknesses as
compared with our approach.

As we have shown in Section \ref{sec:pbqp} the primitive selection problem can
be transformed into the NP-hard PBQP problem.  However, it is worth noting that
the primitive selection problem is NP-hard only in the presence of multiple
data layouts for tensors in the network. If a fixed canonical data
layout is selected for all layers, the problem becomes much easier.

If the NP-hardness of primitive selection can be eliminated by keeping all data
in a canonical layout, then would it not be easier to simply use canonical data
types rather than solve a difficult selection problem? Our experiments show
that this approach can yield high performance in some situations (see VGG-C in
Figure~\ref{fig:results-x86-single-threaded}), but it is always outperformed by
the optimal selection. In some cases, the gap is very wide (see AlexNet in
Figure~\ref{fig:results-x86-single-threaded}).

Our experimental data shows that even simple variations on input data layout
can have a significant impact on data locality, and therefore upon the
execution time of DNN primitives. By solving the primitive selection problem
using our PBQP formulation, we can find an optimal solution that takes data
layout transformation costs into account.

\paragraph{Heuristics}

A similar question arises around other heuristics that might be used
for primitive selection. In other words, is our PBQP formulation a
more sophisticated solution than is needed to solve the primitive
selection problems in our experiments?

There are simpler heuristics (some of which we employ in benchmarking, such as
locally optimal method selection) that still provide good results for the
experiments in Section~\ref{sec:experiments}.

However, the problem with heuristic solutions is not that they perform poorly
on the experimental data that was used during their development. Simple
survivorship bias guarantees that heuristics that do poorly on test data are
quickly dropped by those developing them.

The problem with heuristics is that one is never sure how they will do on
unseen data that arises in practice. We therefore argue that it is worthwhile
using a more sophisticated approach to the problem.

\section{Related Work}
\label{sec:related}

PBQP is an extension of the quadratic assignment problem (QAP)
\cite{Koopmans:1957}, one of the fundamental combinatorial optimization
problems. QAP is NP-hard, as is PBQP, but heuristics have been identified that
can often solve or find approximate solutions for practical instances of PBQP
quickly \cite{Eckstein2003}. PBQP has been used to solve several problems in
compiler optimization such as register allocation for architectures with
irregular instruction sets \cite{Scholz:2002}, and code selection on SSA graphs
\cite{Eckstein2003}.

Latte \cite{Truong:2016} is a domain-specific language, compiler and
runtime for specifying DNNs. Latte provides abstractions such as
neurons, ensembles and connections that allow a programmer to specify
the elements of a DNN. A sophisticated compiler generates optimized
code, and pattern-matches for code regions that can be replaced by
calls to optimized libraries such as GEMM. An advantage of Latte is
that it can automatically fuse successive layers such as convolution
and activation. In contrast, we use a library of highly-optimized
routines, and focus on selecting the best among them rather than
attempting to generate the best code from a high-level specification.

Moskewicz et al. \cite{Moskewicz:2017} proposed Boda, a program
generator for deep neural networks. Boda generates large numbers of
variants of loop nests to implement DNN layers, and uses auto-tuning
to select from among them. Our approach has two significant advantages
when compared with auto-tuning. First, our approach requires much less
time than auto-tuning. Our profiling step requires that we time each
of our methods for each layer in the DNN, whereas auto-tuning
typically requires large numbers of measurements in context. Second,
insofar as the profiled times of each routine are accurate, and the
PBQP instance can be solved in reasonable time, our approach will
provide an optimal selection of layers. A viable future approach might
be to use code generators and auto-tuners to generate the code and
data layouts for given layers and use our approach to combine these
code segments to implement an entire DNN.

GPU vendors and research groups have developed libraries of optimized
routines to implement each layer of DNNs. Perhaps the fastest library
for NVIDIA GPUs is cuDNN \cite{CUDNN:2017}. The cuDNN library provides
several implementations of DNN convolution.

A function is provided to query the library at runtime to heuristically select
the method to implement a given layer given the input parameters. In contrast,
we solve for the global optimal layer selection, taking into account data
layout transformation costs. We also measure profile layer execution times
rather than relying on estimated values. The cuDNN approach is equivalent to
applying our \emph{local optimal} heuristic without fixing the data layout.

TensorFlow XLA is a domain-specific compiler for linear algebra
computations which has been applied to DNNs \cite{Dean:2017}. Compared to simply
invoking primitive functions from a high-level deep learning
framework, XLA allows a stand-alone C/C++ code base to be generated to
implement a DNN. XLA can eliminate intermediate storage buffers and
fuse adjacent layers.

The XLA ahead-of-time compiler is arguably similar
to our own program generation framework which uses C++ templates to
avoid many of the costs of cross-layer transitions and to give the C++
compiler the best opportunity for code optimization. Our layer
selection approach is well-suited to systems such as XLA that generate
DNN code ahead of time.

\section{Conclusion \& Future Work}
\label{sec:conclusion}

Using a PBQP solver in conjunction with per-layer profiling to create a cost
model is an extremely effective tactic for DNN implementation, even under
simplifying assumptions where only convolutional layers are modelled in the
PBQP formulation. We have demonstrated significant gains in DNN inference
performance on both a high performance and embedded processors.

\paragraph{Future Work}

Modelling our problem as an instance of PBQP gives us a great deal of
extensibility. For example, given some
convolution routines which leverage sparsity in the kernel (for example
routines based on a sparse GEMM), our approach can be used to decide whether a
dense or a sparse implementation (and moreover, \emph{which} sparse
implementation) will be faster for any given convolutional layer, with the
addition of a kernel sparsity ratio parameter to the formulation.

Our approach can enable the construction of DNNs using convolution routines
from \emph{different} libraries, if at least one edge in the DT graph connects
a convolution from library A to one from library B. Investigation of the
performance of these ensembles is an exciting prospect for future work.

Our formulation, as mentioned in Section~\ref{sec:pbqp}, does not currently
consider minibatch parallelism, but this can be encoded with another integer
parameter to the model (the minibatch size). This would enable our optimization
approach to select either parallel GEMM or minibatch parallelism on a per-layer
basis.

\section*{Acknowledgment}

\noindent This work was supported by Science Foundation Ireland
grant 12/IA/1381. This project has received funding from the European Union's
Horizon 2020 research and innovation programme under grant agreement No 732204
(Bonseyes). This work is supported by the Swiss State Secretariat for
Education, Research and Innovation (SERI) under contract number 16.0159. The
opinions expressed and arguments employed herein do not necessarily reflect the
official views of these funding bodies. This work was supported in part by
Science Foundation Ireland grant 13/RC/2094 to Lero --- the Irish Software
Research Centre (www.lero.ie).

\bibliographystyle{ACM-Reference-Format}
\bibliography{paper}

\end{document}